\documentclass[aps,prl,twocolumn,groupedaddress,showpacs]{revtex4}

\usepackage{graphicx}

\begin{document}

\title{Novel organic-inorganic layered oxide with spin ladder
structure}

\author{B. Ingham$^1$, J. L. Tallon$^{1,2}$, S. V. Chong$^2$, R.-S. Liu$^{3}$ and L.-Y. Jang$^{4}$}

\affiliation{$^1$Victoria University of Wellington, P.O. Box 600,
Wellington, New Zealand} \affiliation{$^2$Industrial Research
Ltd., P.O. Box 31310, Lower Hutt, New Zealand}
\affiliation{$^3$Department of Chemistry, National Taiwan
University, Taipei, Taiwan, ROC} \affiliation{$^4$National
Synchrotron Radiation Research Center, Hsinchu, Taiwan, ROC}

\date{\today}

\begin{abstract}

Structural analysis of a layered manganese tungstate diaminoalkane
hybrid series suggests that this compound forms a
spin-$\frac{5}{2}$ spin-ladder structure. We use X-ray
diffraction, electron microscopy and x-ray absorption spectroscopy
results to infer the structure. DC magnetization of the manganese
compound and a possible copper analogue appear to show that the
manganese system behaves like 1-dimensional antiferromagnetic
Heisenberg chains (i.e. $J_{\parallel} \gg J_{\perp})$ while the
copper system is fitted well by a $S = \frac{1}{2}$ spin ladder
model, giving the parameters $g = 1.920 \pm 0.008$,
$J_{\parallel}/k_B = -0.3 \pm 1.4$K; $J_{\perp}/k_B = 213.6 \pm
1.1$K. The ability to tune adjacent inorganic layers of the hybrid
materials by altering the length of the organic `spacer' molecules
gives enormous promise for the use of these materials, especially
when doped, to provide a greater understanding of spin ladder
systems.

\end{abstract}

\pacs{
61.10.Ht, 
61.10.Nz, 
75.75.+a, 
81.07.Pr 
}

\maketitle

\section{Introduction}

Spin ladders conceptually represent a transition from 1D to 2D
magnetism, although they have actually proven to be more complex
and exhibit interesting behavior. Since the discovery of
spin-ladder cuprates in 1991 \cite{hiroi}, there has been much
interest in the phenomena observed in these materials. Spin
ladders with an even number of `legs' are predicted (and observed)
to have a spin gap \cite{bar1,dag2,dag1}. The cuprate spin ladder
SrCuO$_3$, and its extensions, have been heavily studied because
of the potential for superconductivity due to the presence of the
spin gap and their structural similarity to the HTS
cuprates\cite{dag1,dag3}. Despite this, superconductivity has not
yet been observed except in the (Sr,Ca)$_{14}$Cu$_{24}$O$_{41}$
system under high pressure \cite{dag1,jpsj1}.

Besides the cuprates, several other systems are also recognized as
spin ladder compounds, including M$^{2+}$V$_2$O$_5$
\cite{dag1,john1,onoda}, (VO)$_2$P$_2$O$_7$
\cite{dag2,bar1,john2}, and (C$_5$H$_{12}$N)$_2$CuBr$_4$
\cite{watson,456}.

The vanadate and cuprate systems listed above all have spin $S =
\frac{1}{2}$ (V$^{4+}$ and Cu$^{2+}$). As such, the only
calculated models available are for $S = \frac{1}{2}$ systems
\cite{john1}. There are a small number of reports of systems with
$S = 1, \frac{3}{2}, 2$ and $\frac{5}{2}$; all of these involve
metal-organic complexes with organic `rungs' of the ladders
\cite{a1,a2,a3,j1,j2}. To the best of our knowledge, no
oxide-based spin ladders have been reported with $S
> \frac{1}{2}$.

Layered organic-inorganic hybrids, based on tungsten oxide layers
separated by organic diamines of various lengths comprise a
structurally 2-dimensional system where the proximity and hence
the coupling of adjacent layers can be controlled, in principle,
by the length of the organic intercalates \cite{us1,us2,us3}.
Recently, we have succeeded in synthesizing layered
organic-inorganic hybrid materials that incorporate magnetic
transition metal ions within the inorganic layers \cite{us4}. Here
we present the structural findings that lead us to believe that
these compounds are capable of forming spin ladder-like
structures, and present a preliminary study of their magnetic
properties.

\section{Experimental}

Manganese tungstate hybrids, (Mn,W)-DAn, were synthesized by
dissolving tungstic acid, H$_2$WO$_4$, and
$\alpha$,$\omega$-diaminoalkanes (C$_n$, n = 2, 6, 8, 12)
separately in hot aqueous ammonia solution and then adding
together. Nitrogen gas was bubbled at a moderate rate through the
stirred solution as a cold aqueous solution of MnCl$_4$ was
quickly added. A precipitate immediately formed that darkened with
time. The excess solvent was evaporated over a water bath. The use
of nitrogen gas was essential to prevent the formation of MnO$_2$.
The products were filtered and washed with ethanol and dried under
vacuum overnight.

A copper tungstate-1,6-diaminohexane hybrid, (Cu,W)-DA6, could not
be synthesized in the same way as the (Mn,W)-DAn samples because
of the preferential formation of the [Cu(NH$_3$)$_4$]$^{2+}$
complex. Instead, copper tungstate hydrate was soaked in a
solution of 1,6-diaminohexane dissolved in toluene for several
weeks to form the hybrid, which was then filtered and washed with
ethanol and dried under vacuum overnight.

XRD powder spectra were obtained using a Philips PW1700 series
powder diffractometer employing Co K$\alpha$ radiation ($\lambda$
= 1.789 \AA). Transmission electron microscopy (TEM) images and
selected area electron diffraction (SAED) photographs were
obtained using a JEOL 2011 high-resolution instrument with a
LaB$_6$ filament operated at 200 kV. Energy dispersive X-ray
analysis (EDX) was performed using a Leo 440 SEM with an Oxford
ISIS EDX system attached. Infrared spectra were obtained using a
Bomem DA8 FT spectrometer over the range 450--4000 cm$^{-1}$ with
a resolution of 2 cm$^{-1}$, using the KBr pellet method. Several
manganese tungstate hybrid samples were sent to the National
Synchrotron Radiation Research Center in Taiwan for x-ray
absorption near edge spectroscopy (XANES) and extended x-ray
absorption fine structure (EXAFS) measurements. Spectra were
obtained at both the Mn K-edge and W L3-edge and fitted using the
FEFF and FEFFIT analysis programs. DC magnetization measurements
were performed using a Quantum Design MPMS XL \textsc{Squid}
magnetometer at applied fields up to 1 T. Powder samples were
ground and 5--20 mg tightly sealed in a gelatin capsule with
negligible magnetic susceptibility.

\section{Structural determination of (Mn,W)-DAn}

Ideally, to determine the structure of an unknown material, single
crystal x-ray diffraction would be performed on suitably large
($\gtrsim 100 \, \mu$m) crystals. However for the
organic-inorganic hybrids, the maximum crystal size observed is
$\sim 10 \, \mu$m \cite{us4}. As a result, a combination of
methods were used to elucidate a structure.

\begin{figure}
\includegraphics*[width=80mm]{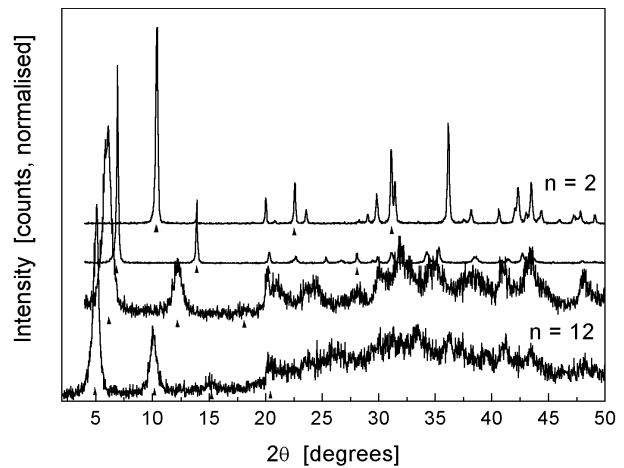}
\caption{\label{fig:xrd} XRD spectra of the (Mn,W)-DAn hybrids, n
= 2, 6, 8, 12 (top to bottom).}
\end{figure}

XRD powder spectra are given for the (Mn,W)-DAn series in Figure
\ref{fig:xrd}. In each case a series of $00 \ell$ lines are
evident (as marked). This is characteristic of layered structures
and the $d$-spacing given by the lines corresponds to the
interlayer spacing (e.g. references \cite{161,190,233}). When
compared with the number of carbons, $n$, in the alkyl chain, the
$d$-spacing follows linearly as $d = 1.05 n + 7.893${\AA}. The
C--C bond length in an alkyl chain is 1.54 {\AA} and the bond
angle is 109.4$^{\circ}$, giving a longitudinal distance per C--C
bond of 1.26 {\AA}. Comparing this with the slope of 1.05, we
conclude that the organic molecules lie at an angle to the organic
plane of 56.5$^{\circ}$.

\begin{figure}
\centerline{\includegraphics*[width=60mm]{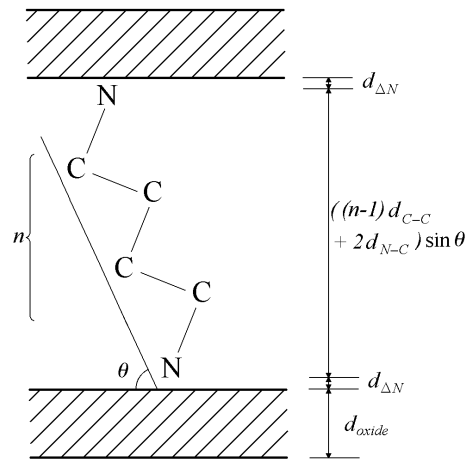}}
\caption{\label{fig:diag} Schematic diagram of the
organic-inorganic hybrid.}
\end{figure}

The distance from one layer to the next is $d = d_{oxide} +
2d_{\Delta N} + ((n-1)d_{C-C} + 2d_{N-C})\sin \theta$, where the
distances $d_i$ are as shown in Figure \ref{fig:diag}.
Extrapolating to $n = 0$ and substituting $d = 7.893$, $d_{N-C} =
1.47 \sin (54.7^{\circ}) = 1.20$, $d_{C-C} = 1.26$ and $\theta =
56.4^{\circ}$, we obtain $d_{inorg} + 2d_{\Delta N} = 6.94${\AA}.
$d_{\Delta N}$ is expected to be very small (i.e. the nitrogen
atoms of the amine groups lying in plane with any apical oxygen
atoms) or even slightly negative. Thus from the oxide layer
thickness of 6.94 {\AA} we conclude there is a bilayer of WO$_6$
octahedra joined at their apices, as one WO$_6$ octahedron is
$\sim$3.65--3.9 {\AA} from one apex to the opposing apex
\cite{304}.

\begin{figure}
\includegraphics*[width=80mm]{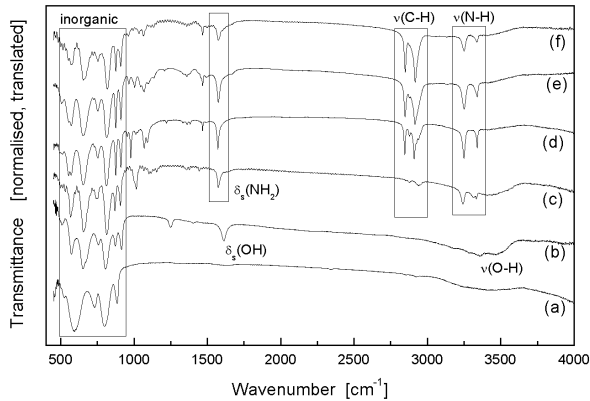}
\caption{\label{fig:ir} Infrared spectra of manganese
tungstate-based materials: (a) MnWO$_4$; (b) MnWO$_4 \cdot
n$H$_2$O; (c) (Mn,W)-DA2; (d) (Mn,W)-DA6; (e) (Mn,W)-DA8; (f)
(Mn,W)-DA12.}
\end{figure}

Infrared spectra of the (Mn,W)-DAn hybrid series (Figure
\ref{fig:ir}) exhibit an identical series of lines at low
wavenumbers, corresponding to the inorganic modes. Hence the
structure of the inorganic layer is the same for all of the
hybrids, regardless of the length of the intercalated organic
molecule. In addition, there are --NH$_2$ modes present and no
--NH$_3^{\phantom{3}+}$ modes, in contrast with the pure tungsten
hybrids (WO$_4$-DAn) \cite{us5}.

\begin{figure}
\includegraphics*[width=80mm]{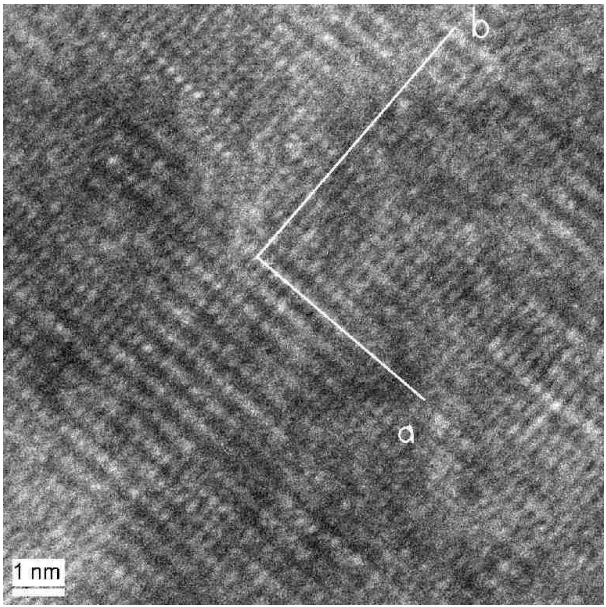}
\caption{\label{fig:tem} High-resolution TEM image of (Mn,W)-DA6,
with axes indicated.}
\end{figure}

\begin{figure}
\includegraphics*[width=80mm]{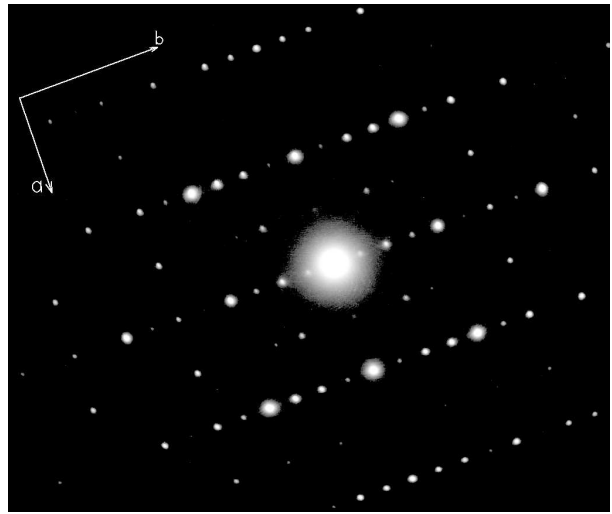}
\caption{\label{fig:saed} Electron diffraction of (Mn,W)-DA6,
showing the $ab$ plane.}
\end{figure}

A high-resolution TEM image of (Mn,W)-DA6 is given in Figure
\ref{fig:tem}. Comparing the spacing of the bright spots
(corresponding to tungsten atoms) along the two principal axes
reveals that the distance between tungsten atoms in the
$b$-direction is almost exactly double that of the $a$-direction.
The manganese atoms are thought to lie in the stripes between the
tungsten atoms. The doubling in the $b$-direction is also evident
from SAED photographs (Figure \ref{fig:saed}), which give $a =
5.03${\AA}, $b = 10.42${\AA} ($\pm 5 \%$); $b/a = 2.070$ and
$\gamma = 89.7^{\circ}$.

From elemental microanalysis combined with EDX we can determine
the composition of the hybrids. The results are given in Table
\ref{tab:composition}. The metal:tungsten:oxygen ratios are also
very close to the inferred stoichiometric values of 1:1:4, even if
the molar ratio of the starting materials was varied to
Mn$_{0.3}$W$_{0.7}$, for instance. No other elements from the
starting materials were detected in the EDX spectra. The organic
ratios are also in good agreement with the expected values: DA2:
C$_2$H$_{8-10}$N$_2$, DA6: C$_6$H$_{16-18}$N$_2$, DA8:
C$_8$H$_{20-22}$N$_2$, DA12: C$_{12}$H$_{28-30}$N$_2$. The
inorganic:organic ratios are relatively consistent and close to
1:0.5, indicating that for every two MnWO$_4$ formula units in the
cell, there is one organic molecule. This is consistent with the
bilayer found via XRD. The uncertainty in H content arises from
its low atomic weight but is pertinent to the question as to
whether the organic molecules are terminated by --NH$_2$ or
--NH$_3^{\phantom{3}+}$. Fortunately, this question is resolved
both by the IR spectra and the x-ray absorption data, as discussed
next.

\begin{table}
\begin{tabular}{|c|ccc|ccc|c|}
\hline
Hybrid material&\multicolumn{3}{c|}{Inorganic}&\multicolumn{3}{c|}{Organic}&I:O Ratio\\
&M&W&O&C&H&N&\\
\hline
(Mn,W)-DA2&1.059&0.941&3.883&2&10.44&1.84&1:0.40\\
(Mn,W)-DA6 & 1.129&0.871&3.742&6&17.3&1.89&1:0.39\\
(Mn,W)-DA8 & 1.226&0.774&3.549&8&19.41&1.85&1:0.54\\
(Mn,W)-DA12 & 0.897&1.103&4.205&12&27.58&1.93&1:0.53\\
\hline
\end{tabular}
\caption{\label{tab:composition} Elemental analysis results for
transition metal hybrids, showing the inorganic results calculated
from EDX analysis, organic results calculated from elemental
microanalysis, and the inoragnic:organic (I:O) ratio. Calculated
for M + W = 2, C = number of carbon atoms in the alkyl chain.}
\end{table}

XANES at the W L1-edge and the Mn K-edge demonstrated that the
manganese tungstate hybrids all have W and Mn valences of 6+ and
2+ respectively. Combined with the EDX analysis, the inorganic
layer has the formula [Mn$^{2+}$W$^{6+}$O$_4$]$^0$. As a result,
the amine terminations of the organic molecules are also neutral
(as already indicated in the IR spectra by the absence of any
--NH$_3^{\phantom{3}+}$ modes). In contrast, the pure tungsten
hybrids consist of WO$_4^{\phantom{4}2-}$ monolayers which must be
balanced by charged diammoniumalkane
[H$_3$N(CH$_2$)$_n$NH$_3$]$^{2+}$ species \cite{us5}.

Like the IR spectra indicating the structural similarity of the
inorganic layer in each of the hybrids, the EXAFS spectra also
appear virtually identical for the three compounds studied
(MnWO$_4 \cdot n$H$_2$O, (Mn,W)-DA2 and (Mn,W)-DA6). MnWO$_4$ was
used as a starting model, and the final fitted results are given
in Table \ref{tab:exafs}. The co-ordination numbers are quite
uncertain past the third (W--W) shell. This is most likely due to
the presence of several shells in this region, to which fits were
also attempted without success. MnWO$_4$ gives the following
shells: W--W 3.281 {\AA}, W--O 3.359 {\AA}, W--Mn 3.521 {\AA},
W--Mn 3.582 {\AA}, W--O 3.602 {\AA}. It is possible then that the
W--W and W--O shells reported in Table \ref{tab:exafs} have much
higher co-ordination numbers than in the actual system, as the
other W--Mn and W--O shells have not been included in the fit, but
would contribute to the spectra in the same region. It is also
important to remember that EXAFS gives an averaged result.

Two shorter W--O bonds (1.7 {\AA}) and four longer ones (2.1
{\AA}) are observed, which most likely correspond to the normally
distorted WO$_6$ octahedra. In MnWO$_4$ the tungsten oxide
octahedra are edge shared, and the W--W distance is consistent
with both W$< ^O_O >$W linked via the two longer W--O planar
bonds, and for W--O--W linked via the shorter apical bond. The
next W--O shell observed probably corresponds to the distance from
one tungsten atom to the apical oxygen atoms of its neighboring
octahedra.

Fitting to the manganese data was more difficult, indicating that
the positions of the manganese atoms are more disordered than
those of the tungsten. In MnWO$_4$ the MnO$_6$ octahedra are
edge-shared to one another and corner-shared to the tungsten oxide
octahedra they sit between \cite{376,296}. However, it appears in
the hybrids that one of the oxygen atoms is missing from the
MnO$_6$ octahedra, forming a square-based pyramid. The next
nearest atom is another manganese, and there are either one (i.e.
Mn dimers) or two (1D chains). However, only these two shells were
used in the fit and there may be other Mn--Mn bonds relatively
close which would appear in subsequent shells (as well as other
Mn--W and Mn--O neighbors).

\begin{table}
\begin{tabular}{|c|c|c|}
\hline
Shell & Length ({\AA}) & Co-ordination number\\
\hline
W--O & 1.737 $\pm$ 0.021 & 2.11 $\pm$ 0.64\\
W--O & 2.121 $\pm$ 0.022 & 4.82 $\pm$ 1.39\\
W--W & 3.233 $\pm$ 0.024 & 6.63 $\pm$ 3.58\\
W--O & 3.337 $\pm$ 0.033 & 12.21 $\pm$ 3.61\\
\hline
Mn--O & 2.124 $\pm$ 0.007 & 4.76 $\pm$ 0.40\\
Mn--Mn & 3.409 $\pm$ 0.041 & 1.28 $\pm$ 0.80\\
\hline
\end{tabular}
\caption{Averaged EXAFS results from fitting analysis of manganese
tungstate hybrids.} \label{tab:exafs}
\end{table}

\begin{figure}
\centerline{\includegraphics*[width=60mm]{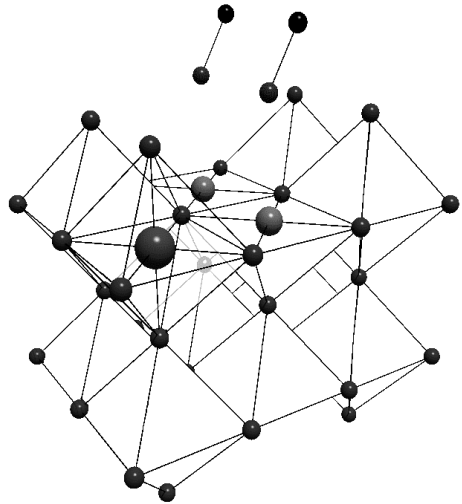}}
\caption{\label{fig:prop} Proposed structure of the inorganic
layer of (Mn,W)-DAn hybrids. The nitrogen and first carbon atom of
the organic molecules are shown. Small spheres = oxygen, middle
spheres = manganese, large spheres = tungsten.}
\end{figure}

Combining the results presented so far, Figure \ref{fig:prop}
shows the proposed structure for the inorganic layer. Here the
organic molecules are positioned in a row above the MnO$_5$ square
pyramids only and not above the WO$_6$ octahedra. There are
therefore two metal cations (W + Mn) to every organic molecule,
consistent with the elemental analysis. Bond-valence sums
\cite{bvs1,bvs2} for this structure, with bond lengths taken from
the EXAFS data, yield valences of $V_{Mn} = 2.03 \pm 0.03$ and
$V_W = 5.56 \pm 0.33$ which are in good agreement with the values
obtained from the XANES results. From this the manganese is
determined to have $S = \frac{5}{2}$ and tungsten has $S = 0$.
Thus any magnetism observed is due solely to the arrangement and
interaction of the manganese atoms.

\section{Magnetization study}

\begin{figure}
\includegraphics*[width=80mm]{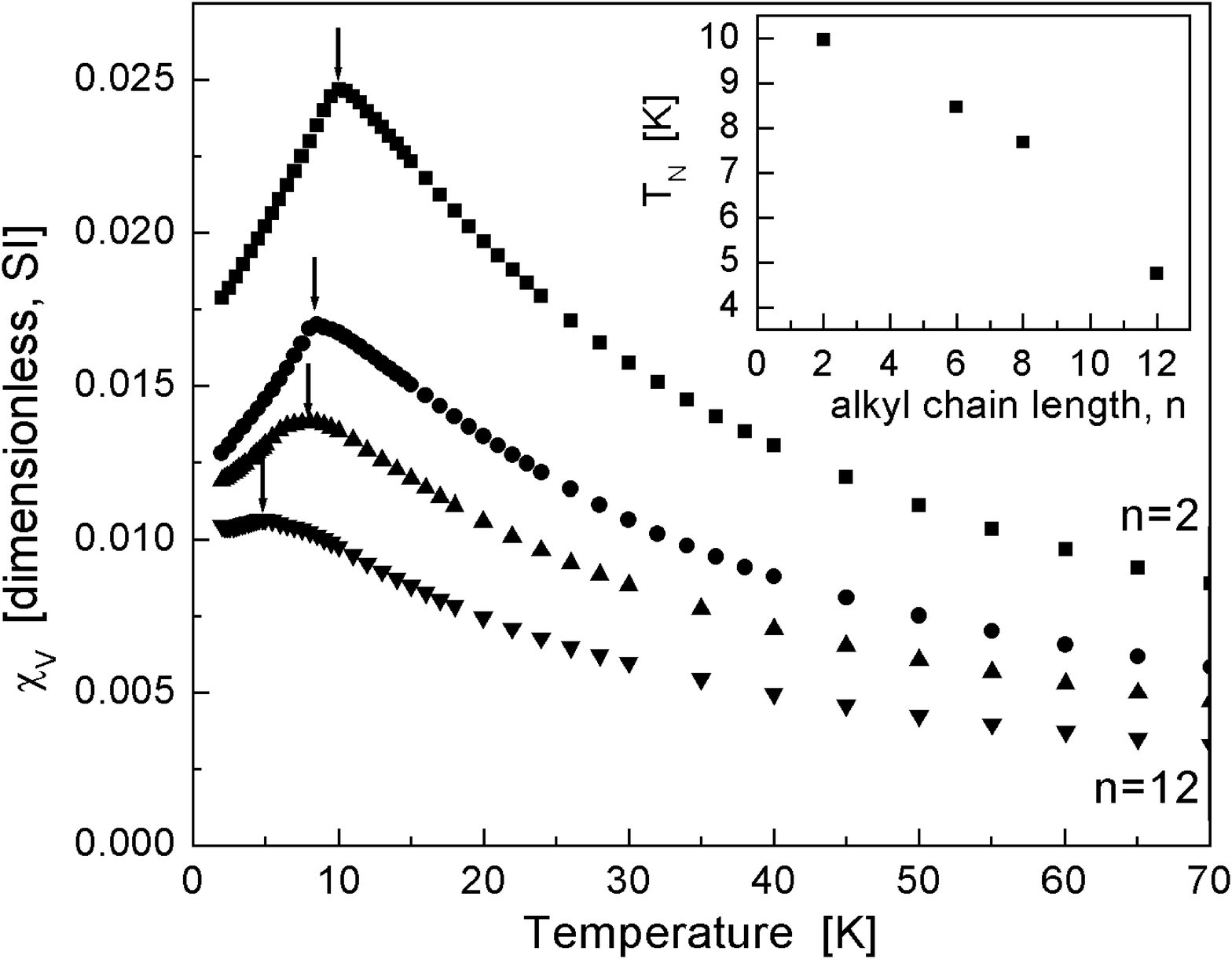} \caption{DC susceptibility of
(Mn,W)-DAn series (n = 2, 6, 8, 12 from top to bottom). $T_N$ is
indicated by the arrows and plotted in the insert versus alkyl
chain length.} \label{fig:Mn-chi}
\end{figure}

The DC susceptibility of a series of (Mn,W)-DAn hybrids is shown
in Figure \ref{fig:Mn-chi}. The field- and zero-field-cooled
curves are identical, indicating that the magnetization is
reversible. There is no field-dependence of the susceptibility
($\propto M/H$) for applied fields up to 1 T.

In each case there is a transition at low temperatures, indicated
by the arrows. For (Mn,W)-DA2 and (Mn,W)-DA6 especially, the
feature is cusp-like, characteristic of an antiferromagnetic (AF)
ordering transition. (The clarity of the transition in these two
samples is probably related to their increased crystallinity
compared with the longer chain alkane hybrids.)

The N\'{e}el temperature is plotted as a function of alkyl chain
length in the insert of Figure \ref{fig:Mn-chi}. This shows a
monotonic decrease in $T_N$ as the alkyl length, and hence the
interlayer spacing, increases. It is expected that as the
interlayer spacing is increased the interlayer coupling will
decrease, which in turn will cause the transition temperature to
decrease. This is qualitatively what is observed.

Curie-Weiss fits to the high-temperature parts of the curves
($T>50$K) display consistent Weiss temperatures: (Mn,W)-DA2:
-19.6K, (Mn,W)-DA6: -19.6K, (Mn,W)-DA8: -19.9K, (Mn,W)-DA12: -20.1
K. This is in contrast with the `parent' MnWO$_4$ compound which
has a Weiss temperature of -72 K \cite{440}. This parent compound
is not layered and hence exhibits stronger AF ordering in 3
dimensions \cite{296}, which is weakened in the case of the
layered hybrid materials.

\begin{figure}
\includegraphics*[width=80mm]{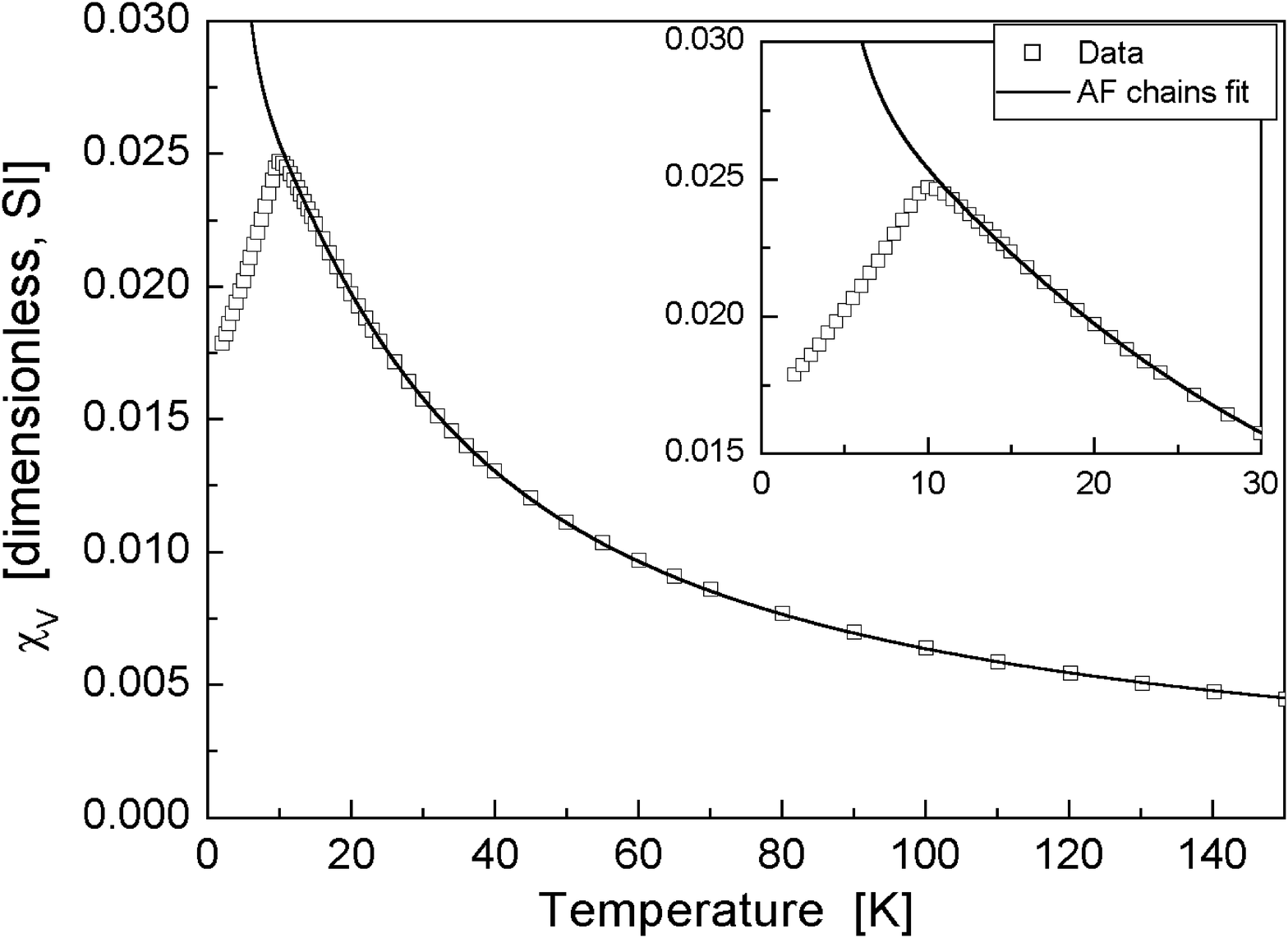} \caption{DC
susceptibility of (Mn,W)-DA2 with best fit curve to the Heisenberg
AF chain model (fitted to $T>T_N$).} \label{fig:Mn-fit}
\end{figure}

From the proposed structure in Figure \ref{fig:prop}, as
determined from TEM, EXAFS, and XRD, the arrangement of the
manganese ions resembles that of a spin ladder compound. However,
no empirical or theoretical model yet exists for $S > \frac{1}{2}$
and in the few literature examples of spin ladder compounds with
$S > \frac{1}{2}$, comparisons are made between the observed
magnetization and models for chains and dimers \cite{a1,a2}. The
(Mn,W)-DA2 DC susceptibility is shown in Figure \ref{fig:Mn-fit}
to a model for antiferromagnetic Heisenberg $S = \frac{5}{2}$
chains \cite{dingle,fisher}, given by

\begin{eqnarray}
\chi = \frac{C}{T - \Theta} \left( \frac{1-u}{1+u} \right) + TIP, \nonumber \\
\nonumber \\
u = \frac{T}{T_0} - \coth \frac{T_0}{T}, \qquad T_0 =
\frac{2JS(S+1)}{k_B} \nonumber
\end{eqnarray}

This model resembles the Curie-Weiss law, with an additional
temperature-dependent perturbation, where $J$ is the coupling
within the chains and $S = \frac{5}{2}$ is the spin of the ion.
(This is the limit of the spin ladder as $J_{\perp} \rightarrow
0$.) The chi-squared degree of fit for the AF chain model
indicates a better fit than the simple Curie-Weiss, and matches
the curve almost up to the AF transition (as shown in the insert).

From the fit parameters, the intrachain coupling value, $J$, has a
value of $J/k_B = -1.20$ K. This is comparable with similar values
obtained for layered manganese phosphate hybrids of $-0.79$ K
\cite{352} and $-1.66$ K \cite{349}.

\section{Copper analogue}

If a compound with the same structure as Figure \ref{fig:prop}
could be formed with copper instead of manganese, this would
represent a new variety of copper oxide $S = \frac{1}{2}$ spin
ladder, with the potential for doping in order to obtain new
properties.

\begin{figure}
\includegraphics*[width=80mm]{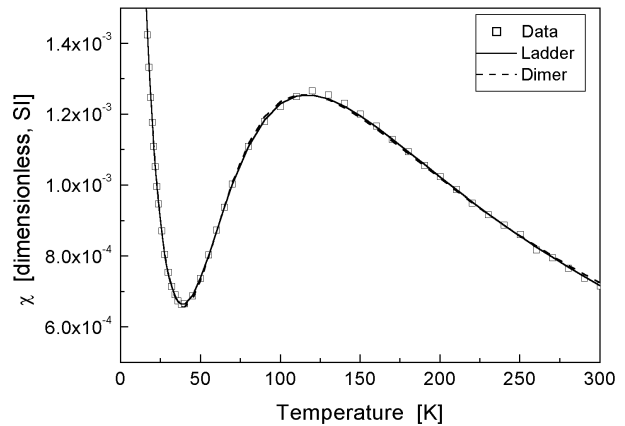} \caption {DC
susceptibility of (Cu,W)-DA6 with best fit curves for the spin
dimer and spin ladder models.} \label{fig:Cu-DA6}
\end{figure}

A copper tungstate-DA6 hybrid has been synthesized, however like
the manganese sample, full structural analysis has not been
possible because of the small crystallite sizes. Despite this,
there is reason to believe it may form a related structure, as its
magnetization curves strongly resemble those observed in other
spin-$\frac{1}{2}$ ladder systems. Figure \ref{fig:Cu-DA6} shows
the DC susceptibility of (Cu,W)-DA6. There is no irrevsibility and
the susceptibility is independent of field.

A dimer model (Equation \ref{eq:Cu-dimer}) suggested in Ref.
\cite{323} is as follows:

\begin{equation}
\chi = \frac{2Ng^2\mu_B^{\phantom{B}2}}{3k_BT} \left[
1+\frac{1}{3}
\exp (-2J/k_BT) \right]^{-1} + \frac{C}{T-\Theta} + TIP \\
\label{eq:Cu-dimer}
\end{equation}

It fits the (Cu,W)-DA6 data extremely well, with the following fit
parameters: $g = 0.413 \pm 0.001$; $J/k_B = -102.9 \pm 0.2$K; $C =
0.0277 \pm 0.0003$K; $\Theta = 0.17 \pm 0.08$K; $TIP = -0.00022
\pm 0.000007 \, (\chi^2/DoF = 4.437 \times 10^{-11})$.

A spin ladder model for $S = \frac{1}{2}$ was also fitted to the
data. Due to the good agreement between the spin dimer model and
the experimental data, we assume that in the spin ladder model
$J_{\perp} \gg J_{\parallel}$ and all other exchanges ($J_{diag}$,
$J_{interchain}$, etc.) are negligible. We used the fit in Ref.
\cite{john1} for isolated chains with $J_{\perp}/J_{\parallel}
\geq 1$:

\begin{eqnarray}
\chi(T) = \frac{N_Ag^2\mu_B^{\phantom{B}2}}{J_{\perp}}
\chi^{\ast}(t) +
\frac{C}{T-\Theta} + TIP; \label{eq:Cu-ladder} \\
\chi\ast(t) =
\frac{\exp(-\Delta_{fit}^{\ast}/t)}{4t}\mathcal{P}^p_q(t);\nonumber \\
t \equiv \frac{k_BT}{J_{\perp}},\nonumber \\
\mathcal{P}^p_q(t) = \frac{1+ \sum_{n=1}^p N_n/t^n}{1+
\sum_{n=1}^q D_n/t^n} \nonumber
\end{eqnarray}

\noindent where the parameters $\Delta_{fit}^{\ast}$, $N_n$ and
$D_n$ (all functions of $J_{\parallel}/J_{\perp}$) are given in
Appendix 5 of Ref. \cite{john1}.

The following fitted parameters were obtained:

$g = 1.920 \pm 0.008$; $J_{\parallel}/k_B = -0.3 \pm 1.4$K;
$J_{\perp}/k_B = 213.6 \pm 1.1$K; $C = 0.0268 \pm 0.0003$K;
$\Theta = -0.0015 \pm 0.0004$K; $TIP = -0.0003 \pm 0.00001 \,
(\chi^2/DoF = 2.3718 \times 10^{-11})$.

The two fits are compared in Figure \ref{fig:Cu-DA6} with the
experimental data. The ladder model is shown as the solid line and
the dimer model as the dotted line. Comparing the two fits, we
note firstly that the assumption $J_{\perp} \gg J_{\parallel}$ is
valid and therefore the spin dimer model should be sufficient to
describe the (Cu,W)-DA6 system. However, the spin dimer model does
not give an accurate $g$ factor - the expected value is $\sim 2$,
whereas the dimer model yields $g=0.413$. However the spin ladder
model gives $g=1.920$, in good agreement with the expected value
and consistent with other Cu$^{2+}$ $S=\frac{1}{2}$ systems
\cite{323}. The exchange constant $J_{\perp}$ between the two Cu
atoms of the dimer in the ladder model is roughly double that
obtained for the dimer model; this is due to a difference of
notation in the two models. (In the exponential term of the dimer
model (Equation \ref{eq:Cu-dimer}), one has a factor of 2 that is
not present in the exponential term of the spin ladder fit
(Equation \ref{eq:Cu-ladder})). The values of $C$ and $TIP$ are
virtually identical. While $\Theta$ has different signs in each of
the two models, it has a small value so can be essentially treated
as zero.

\section{Conclusions}

Indirect methods of structural analysis of a manganese tungstate
diaminoalkane hybrid series strongly suggest that this compound
forms a spin ladder-type structure with $S = \frac{5}{2}$.
Magnetization results indicate that the system behaves like a 1D
AF Heisenberg chain, so it is likely that $J_{\parallel} \gg
J_{\perp}$. A copper tungstate hybrid on the other hand fits well
to a spin ladder model with $J_{\perp} \gg J_{\parallel}$, thus
behaving more like a spin dimer system. The fact that the spin
ladder model gives a more reasonable value of $g$, and has a
better $\chi^2/DoF$ value than the dimer model, indicates that the
spin ladder model probably has some validity.

Future work in this area includes attempts to dope the inorganic
layer through electrochemical means. It should also be possible to
form samples with composition Mn$_{1-x}$Cu$_x$WO$_4 \cdot$DAn to
determine the structure and magnetism across the solid-solubility
series.

\section{Acknowledgements}

The authors would like to acknowledge the financial assistance
from the New Zealand Foundation of Research Science and Technology
(Contract number: IRLX0201), The Royal Society of New Zealand
Marsden Fund, and the MacDiarmid Institute for Advanced Materials
and Nanotechnology (Victoria University, New Zealand).

\bibliography{spinlad}
\end{document}